# Data-Driven Model for Elastomers under Simultaneous Thermal and Radiation Exposure

Pouyan Nasiri[a], Leonard S. Fifield[b], Hadis Nouri[c], Roozbeh Dargazany[a,*]

[a]*Department of Civil and Environmental Engineering, Michigan State University*
[b]*Pacific Northwest National Laboratory, Richland, WA, USA*
[c]*Karax LLC, East Lansing, MI, USA*

**Abstract**

We present a physics-informed neural network (PINN) framework for predicting the mechanical performance of elastomers exposed to concurrent thermal and gamma-radiation exposure, such as elastomers in nuclear cables or space electronics. Our demonstrated approach integrates the dual-network hypothesis with the microsphere concept to represent soft and brittle sub-networks, while embedding physical laws directly into the machine learning process. Hard constraints (e.g., incompressibility, bounded network fractions) are enforced through network architecture, and soft constraints (e.g., monotonicity, polyconvexity, and fading effects) are imposed through the loss function. This integration reduces the effective search space, guiding the optimization toward physically admissible solutions and enhancing robustness under sparse data. Validation against published datasets on silicone rubber, ethylene propylene diene monomer, and silica-reinforced silicone foam shows accurate predictions of stress–strain behavior and elongation-at-break at exposure times not used for training. Results confirm that physics-informed constraints improve extrapolation, capture synergistic thermal–radiation effects, and provide a reliable tool for lifetime assessment of nuclear cable insulation and other radiation-exposed elastomers.

## 1. Introduction

Elastomers have been extensively used in various industrial applications. The environmental conditions can significantly affect the durability of elastomers. The main contributors to elastomer degradation include temperature, radiation, and humidity. For useful aging management, it is imperative to predict the performance of these materials under external deteriorating conditions. Here, our focus is to predict the durability of elastomers under exposure to simultaneous combined thermal and radiation. As an application, the cables of nuclear power plants are manufac-tured from elastomers, and these materials are exposed to high temperatures and high-energy radiation. Under these extreme conditions, they should endure for years.

*Corresponding author: roozbeh@msu.edu

Many studies have been published on the experimental characterization of elastomers following thermal and radiation degradation [1, 2, 3, 4, 5, 6, 7, 8, 9, 10]. In [11], Shimida et al. investigated the degradation of silicon rubber employed in the cables of nuclear power plants. They observed the changes in the mechanical properties of silicon rubber, such as elongation at break and tensile strength, under sequential and simultaneous thermal and radiation exposure. In sequential aging, thermal or radiation exposure occurs first, followed by the other exposure. On the other hand, simultaneous aging is defined as the material being exposed to both thermal and radiation concurrently. As one of their main conclusions, the degradation level of simultaneous aging is considerably higher than that of sequential aging. In another important study [12], Gillen and Clough proposed an empirical method called time-temperature-dose rate superposition for describing combined thermal and radiation environments. Their method is an extension of the time-temperature superposition technique for polymer aging. The goal is to predict the service life of polymers using accelerated experiments. In a recent study, Celina et al. [13] proposed multiple empirical models for combined thermal-radiation degradation from the basic to the most comprehensive. Time- and dose-to-equivalent damages were defined based on the rate of degradation used in their studies.

Various approaches have been proposed to apply the constitutive model for elastomers. The two predominant frameworks are phenomenological and micromechanical. Often, the physical parameters of phenomenological methods do not have a physical basis, such as with the Mooney-Rivlin model [14] or the Ogden model [15]. However, micromechanical approaches are based on physical parameters such as elastomer crosslink density or chain length [16, 17]. In addition, models may be formulated based on strain variants, stretch ratios, or a mixed formulation. A general review paper on the constitutive models for elastomers can be found in [18]. Mohammadi et al. [19, 20] proposed a novel approach towards modeling combined thermal and radiation aging of elastomers. They employed the microsphere concept with the dual network assumption [21] to derive the constitutive model. Maiti et al. [22] proposed a constitutive model for elastomers under radiation aging, combining the Ogden model and dual network hypothesis.

Data-driven models have increasingly been applied to characterize material behavior [23, 24, 25, 26, 27, 28]. Ghaderi et al. [27] proposed a constitutive framework to predict the quasi-static response of elastomers by combining machine learning approaches with the microsphere concept. They demonstrated both the simplicity and accuracy of their model through comparisons with phenomenological and micromechanical formulations. Subsequently, they extended their framework to capture thermo-oxidative aging of elastomers [29], embedding aging time and temperature into the model and validating it against extensive experimental data, including intermittent and relaxation tests at multiple conditions. More recently, however, it was shown that a single network is insufficient to reproduce the stress-stiffening behavior of elastomers when both intermittent and relaxation data are considered [30]. Our proposed solution is based on the dual-network hypothesis by Tobolsky et al. [21], which is further advanced by network evolution theory and infused into a feedforward neural network.

In this work, we introduce a novel data-driven paradigm to address the simultaneous effects of thermal and radiation on elastomer degradation. Our approach integrates deep neural networks with the microsphere concept to derive a constitutive model for elastomers exposed to extreme environments. Unlike earlier studies that relied on a single-network framework, our formulation is rooted in the dual-network hypothesis, which is essential for capturing the complex behavior of elastomers. Furthermore, while previous data-driven studies have rarely enforced physical constraints, we explicitly incorporate such constraints into both the decay and evolution components of our model. In addition, we investigate thermodynamic consistency and frame indifference for our model. Furthermore, we impose polyconvexity along with growth conditions on our model by construction to ensure mathematical consistency.

The remainder of this paper is structured as follows. Section 2 describes the experimental setup. Section 3 introduces the dual-network and microsphere concepts. Section 4 presents the proposed data-driven constitutive model for elastomers under combined thermal and radiation exposures, with emphasis on the incorporation of physical constraints. Section 5 develops a failure prediction framework based on elongation at break, and Section 6 validates the model against experimental data. Finally, Section 7 summarizes the main conclusions.

## 2. Experimental Method

### 2.1. End-of-Life Parameters

To consider long-term reliability and performance in safety-related functions, several measurable parameters are employed to evaluate the aging state and define the end-of-life (EOL) of elastomers. The most relevant include mass change, elongation at break, carbonyl index, total color difference, and indenter modulus.

- **Mass change** reflects the uptake or loss of volatile compounds, plasticizers, or absorbed chemicals during aging. Significant mass loss typically indicates evaporation or leaching of additives, whereas mass gain suggests oxidation of the polymer hydrocarbon chain. Both conditions may alter mechanical performance. Thresholds of significant change depend on material formulation, but deviations beyond ±5% of the original mass are often regarded as notable indicators of aging.

- **Elongation at break (EAB)** is a critical mechanical property reflecting flexibility and ductility. A reduction of tensile EAB to below 50% of its original value is widely adopted as a practical EOL criterion, as it denotes embrittlement and increased cracking risk under mechanical stress.

- *The carbonyl index (CI)* obtained via Fourier-transform infrared spectroscopy (FTIR), quantifies oxidation by tracking the growth of carbonyl functional groups. An increasing index reflects chain scission and oxidative degradation. Although thresholds are material-specific, a sharp increase compared to unaged samples signifies chemical aging.

- *Total color difference ($\Delta E$)* serves as a non-destructive visual indicator of surface degradation, measured spectrophotometrically. High $\Delta E$ values may reflect oxidation of polymers, degradation of additives in the polymer formulation, or other chemical changes associated with aging.

- *Indenter modulus (IM)* characterizes surface stiffness through non-destructive mechanical testing. A significant increase relative to the baseline indicates hardening and loss of elasticity, features commonly associated with aging in cables. It is especially valuable for in-situ monitoring where components cannot be removed from service.

Collectively, these parameters provide a comprehensive picture of elastomer aging and support decision-making in aging management programs and life-extension strategies, particularly in safety-critical applications in nuclear facilities. Table 1 shows the thresholds of EOL parameters. In addition, Figure 1 illustrates the evolution of EOL parameters during aging.

| Parameter | Typical EOL Threshold |
| --- | --- |
| **Mass Change** | ±2–5% mass variation; ±1–2% for high-reliability applications |
| **Elongation at Break** | 50% of original value (e.g., from 300% to 150%) |
| **Carbonyl Index** | 2–5× increase from initial value |
| **Color Difference** | $\Delta E > 5$: Significant visual change |
| **Indenter Modulus** | ±20% change |

Table 1: Summary of end-of-life thresholds for common elastomer degradation indicators.

*2.2. Simultaneous versus Sequential Aging*

Elastomeric materials, widely used in radiation-prone and high-temperature environments, degrade over time under external stressors such as gamma radiation and thermal exposure. When these stressors are applied in succession (sequential aging), their effects can often be treated independently. For example, an elastomer may first be exposed to gamma irradiation and later subjected to elevated temperatures during service. In such cases, degradation is gen-erally additive, with the final properties reflecting the cumulative impact of each stressor. This enables relatively straightforward assessments and modeling using independent exposure data.

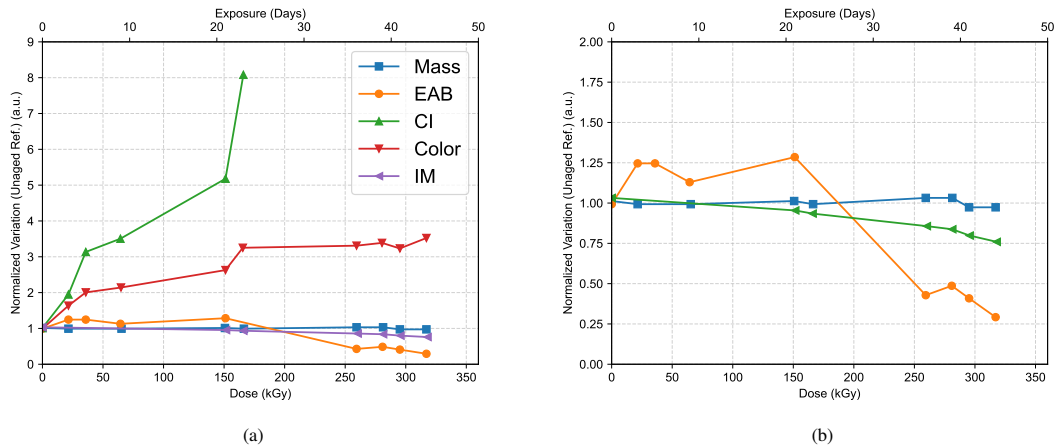

(a)                          (b)

Figure 1: Typical progression of the end-of-life parameters for elastomers (cross-linked polyethylene) during simultaneous thermal and radiation aging (a) considering all parameters (b) considering only mass, EAB, and IM. [31]

In contrast, simultaneous exposure to gamma radiation and heat produces synergistic aging, characterized by complex and accelerated degradation pathways. Gamma irradiation generates free radicals within the polymer matrix, while elevated temperatures enhance radical mobility and reactivity. This intensifies chain scission, crosslinking, and oxidative degradation. These coupled processes interact in a nonlinear fashion, sometimes producing damage levels greater than the sum of individual exposures. Relying solely on sequential aging data may therefore underestimate material degradation, leading to unexpected failures. Accurate lifetime prediction in environments such as nuclear reactors or space applications requires explicit consideration of these nonlinear synergistic effects. A comparison between sequential versus simultaneous aging is outlined in Figure 2. In this study, the mechanical behavior of aged elastomers is ultimately characterized by tensile testing at room temperature.

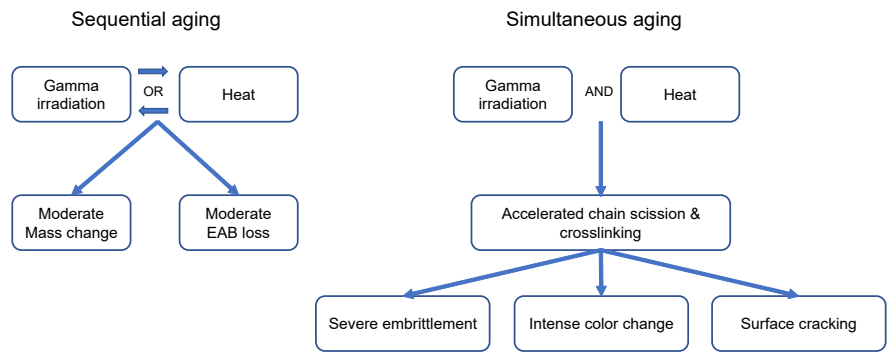

Figure 2: Comparison between sequential versus simultaneous aging: In sequential aging, the material is first exposed to Gamma or heat, followed by the other one, with moderate damage, while simultaneous aging happens during simultaneous exposure of gamma radiation and heat, sometimes with severe damage. Note that this is only true for some materials and exposure conditions, not a general fact.

## 2.3. Micro- & Macro-Scale Aging

At the microscale, simultaneous exposure to gamma irradiation and elevated temperature induces physicochemical transformations that compromise structural integrity. Gamma irradiation ionizes polymer chains, creating free radicals that initiate degradation. Two competing processes dominate: chain scission, which breaks chains into shorter segments, reducing molecular weight and leading to softening and reduced tensile strength; and crosslinking, which forms covalent bonds between chains, producing a denser, stiffer network that increases brittleness. Elevated temperature accelerates both reactions by increasing radical mobility and reactivity. Oxygen plays a crucial role, as higher temperatures enhance diffusion into the polymer, where oxygen reacts with radicals to form peroxyl species. These propagate oxidative reactions, generating polar functional groups (e.g., carbonyls, carboxylic acids) and producing chemically weakened regions prone to microcracks and crazes. The balance between chain scission and crosslinking is strongly dependent on polymer type, dose rate, thermal conditions, and oxygen availability. Figure 3 schematically illustrates these competing reactions for cross-linked polyethylene (XLPE).

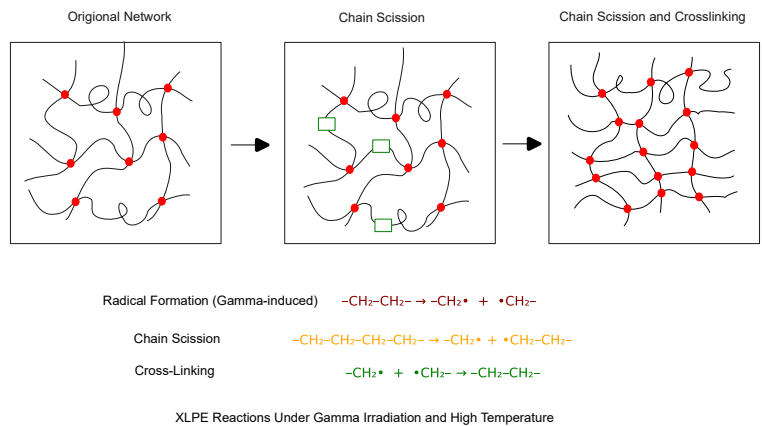

Figure 3: Schematic illustration of network alteration in elastomers through chain scission and crosslinking. Long, flexible chains progressively transform into shorter, stiffer, and more brittle chains under combined gamma and thermal exposure.

These molecular-level changes ultimately manifest at the macroscale. Common indicators include color change (yellowing or browning), surface cracking, and dimensional variations such as swelling or shrinkage. Mechanical properties also degrade, with reduced elongation at break and diminished elasticity. In electrical applications, dielectric strength is reduced, posing additional reliability concerns. Macro-level manifestations thus provide visible evidence of microscale degradation mechanisms and are essential for correlating laboratory findings with real-world service conditions.

## 2.4. Experimental Data and Discussion

This section summarizes the experimental datasets used for three elastomeric materials: Silicone Rubber (SR), Ethylene Propylene Diene Monomer (EPDM), and Silica Reinforced Silicone Foam (SRSF). The analysis focuses on the constitutive response (stress–strain behavior) and elongation at break (EAB). In general, EAB decreases with increasing aging time, while the constitutive curves of all materials exhibit progressive stiffening.

For SR, data were obtained from Shimida et al. [11]. Specimens of thickness 0.5, 1.0, and 2.0 mm were prepared and exposed to various combinations of temperature, time, and radiation dose rate (1 kGy/h). Figure 4 summarizes the extracted dataset for EAB.

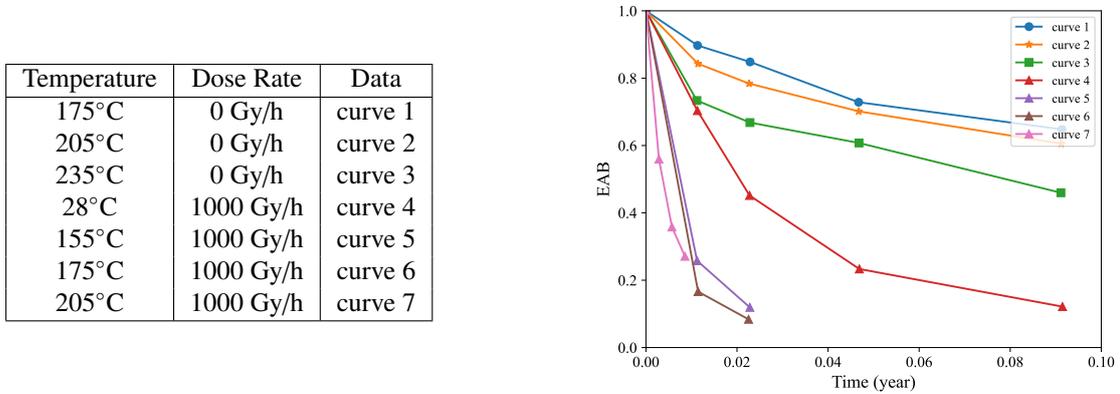

| Temperature | Dose Rate | Data |
|---|---|---|
| 175°C | 0 Gy/h | curve 1 |
| 205°C | 0 Gy/h | curve 2 |
| 235°C | 0 Gy/h | curve 3 |
| 28°C | 1000 Gy/h | curve 4 |
| 155°C | 1000 Gy/h | curve 5 |
| 175°C | 1000 Gy/h | curve 6 |
| 205°C | 1000 Gy/h | curve 7 |

Figure 4: SR dataset for elongation at break [11].

For EPDM, the dataset was extracted from Figure 6 in [32]. Figure 5 shows the derived data points selected at specific temperatures and dose rates. It should be mentioned that the dose rates were directly reported.

For SRSF, data were extracted from [33]. The irradiation dose rate was kept constant at 20 Gy/min (1.2 kGy/h). As summarized in Table 2, only five data points are available for EAB. Since the proposed model for EAB involves five parameters (see Section 5), ideally, more experimental points are required for robust fitting and training. In this work, initial parameter estimates were taken from previous datasets, which enabled tuning despite the limited data.

| Temperature | Dose Rate | Aging Time | EAB |
|---|---|---|---|
| 20°C | 0 Gy/h | 0 h | 100% |
| 20°C | 1200 Gy/h | 83.3 h | 81.4% |
| 20°C | 1200 Gy/h | 166.6 h | 62.8% |
| 20°C | 1200 Gy/h | 333.3 h | 37.1% |
| 20°C | 1200 Gy/h | 500 h | 28.8% |

Table 2: SRSF dataset for elongation at break [33].

| Temperature | Dose Rate | Data    |
|-------------|-----------|---------|
| 55°C        | 106 Gy/h  | curve 1 |
| 70°C        | 106 Gy/h  | curve 2 |
| 85°C        | 106 Gy/h  | curve 3 |
| 40°C        | 1390 Gy/h | curve 4 |
| 70°C        | 1390 Gy/h | curve 5 |
| 70°C        | 2760 Gy/h | curve 6 |

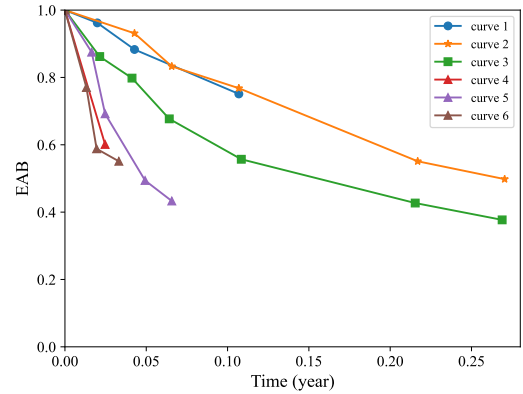

Figure 5: EPDM dataset for elongation at break [32].

## 3. Dual Network Hypothesis & Micro-sphere Concept

We adopt the mathematical notation and framework introduced in our previous study [30]. The dual-network hypothesis, originally proposed by Tobolsky et al. [21], is grounded in two key experimental protocols: stress-relaxation and intermittent tests. In a stress-relaxation test, a specimen is stretched and held at constant temperature while the stress decay is monitored over time. In contrast, during an intermittent test, the specimen is aged in an oven under fixed temperature for a prescribed duration, then removed and mechanically characterized. The differences observed between these two tests demonstrate that both chain scission and crosslinking occur simultaneously during elastomer aging. Depending on environmental conditions, elastomers may exhibit softening, hardening, or a combination of both. This concept is illustrated schematically in Figure 6.

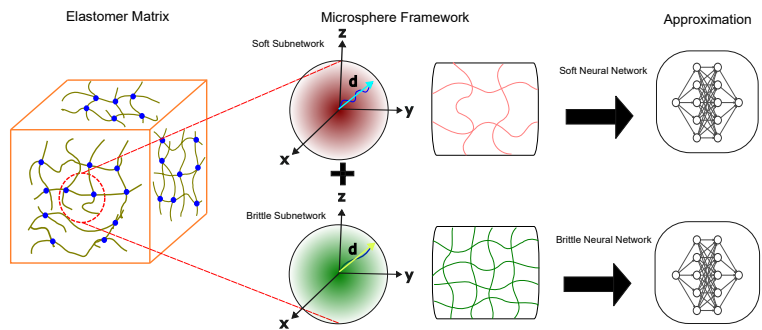

Figure 6: Schematic representation of the dual-network hypothesis and micro-sphere concept.

At the onset of aging, the elastomer consists primarily of a soft network. Over time, this network gradually transforms into a brittle one, and at sufficiently long times only the brittle network remains. According to the dual-network hypothesis, the elastomeric response is described as a fusion of these two networks. Under coupled thermal and radiation exposures, the total strain energy can be expressed as

$$\Psi_M = \rho_s(t, T, \dot{D})\, \Psi_s(\mathbf{F}_s) \;+\; \rho_b(t, T, \dot{D})\, \Psi_b(\mathbf{F}_b), \tag{1}$$

Here, $t$, $T$, and $\dot{D}$ denote time, temperature, and dose rate, respectively; $\rho_s$ and $\rho_b$ represent the contributions of the soft and brittle networks; $\mathbf{F}_s$ and $\mathbf{F}_b$ are the corresponding deformation gradients; $\Psi_M$ is the total network strain energy, and $\Psi_s$, $\Psi_b$ are the strain energy functions of the soft and brittle networks. The weighting functions $\rho_s$ and $\rho_b$ evolve with aging and are interpreted as the decay and growth functions of the elastomer networks.

Considering the multiplicative nature of weighting functions $\rho_s$ and $\rho_b$ makes them act as shape functions between $\Psi_s$ and $\Psi_b$, ensuring that the total network response is always partitioned between the soft and brittle contributions. In an ideal case where the overall integrity of the polymer matrix is preserved during transformation, one can assume $\rho_s + \rho_b = 1$. However, considering mass loss of the network, we introduced

$$\phi := \rho_s + \rho_b, \qquad \phi \leq 1. \tag{2}$$

where $\phi < 1$ incorporates a fading effect, representing irreversible mass loss and polymer chain detachment from the load-bearing network. This refinement captures the physical reality of elastomer degradation, where volatilization, oxidation, and fragmentation reduce the effective number of active chains contributing to the mechanical response.

Several aging models have been proposed to describe degradation, but most rely on Arrhenius-type rate equations. These models often fail to capture nonlinearities at lower temperatures. For example, the inverse temperature effect—where polymer degradation accelerates at lower ambient temperatures for constant dose rates—cannot be described adequately by a conventional Arrhenius framework. This effect is illustrated in Figure 7, where at 50°C there exists an abnormal behavior compared with the room temperature aging. Our model does not account for the inverse temperature effect, which is only observed in some polymers [34]. Long-term studies further demonstrate that many polymers exhibit non-Arrhenius behavior: activation energies estimated from short-term, high-temperature experiments often decrease as the temperature approaches service conditions, highlighting the need for alternative predictive methodologies.

**Microsphere Concept** describes the 3D elastomer network as an assembly of one-dimensional subnetworks, each oriented in a specific direction and sustaining only uniaxial deformation. The strain energy of each network (soft or brittle) is obtained by integrating the directional contributions over the unit sphere:

$$\Psi_* = \frac{1}{A_*} \int_S \Psi_*^d \, da^d, \qquad * \in \{s, b\}, \tag{3}$$

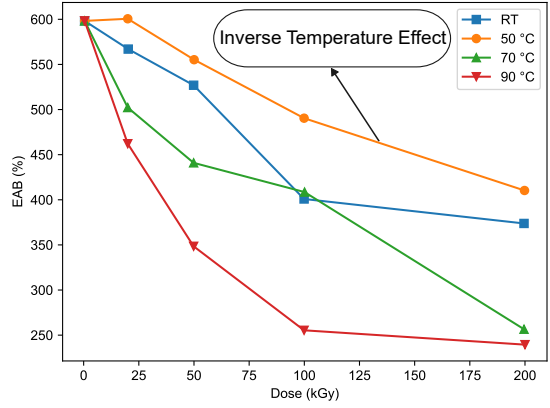

Figure 7: Typical inverse temperature effect taken from [34].

where $A_*$ is the microsphere area for the given network, $\Psi_*^d$ is the directional strain energy of a subnetwork oriented along $\boldsymbol{d}$, and $a^d$ is the associated surface element.

## 4. Constitutive Modeling

We now present the constitutive model for elastomers subjected to combined thermal and radiation environments. The objective is to predict the stress–stretch relation under extreme conditions. Starting from the total strain energy, the first Piola–Kirchhoff stress is given by

$$\mathbf{P} = \frac{\partial \Psi_M}{\partial \mathbf{F}} - p\,\mathbf{F}^{-T}, \qquad (4)$$

where $p$ enforces incompressibility. Using the chain rule, the derivative of the strain energy with respect to the deformation gradient becomes

$$\frac{\partial \Psi_*}{\partial \mathbf{F}_*} = \frac{\partial \Psi_*}{\partial \lambda_*^d}\frac{\partial \lambda_*^d}{\partial \mathbf{F}_*}, \qquad * \in \{s,b\}, \qquad (5)$$

with $\lambda_*^d = \sqrt{\boldsymbol{d}\mathbf{C}_*\boldsymbol{d}}$ the stretch ratio along $\boldsymbol{d}$ and $\mathbf{C} = \mathbf{F}^T\mathbf{F}$ the right Cauchy–Green tensor. The derivative of the stretch ratio is expressed as

$$\frac{\partial \lambda_*^d}{\partial \mathbf{F}_*} = \frac{1}{\lambda_*^d}\mathbf{F}_*(\boldsymbol{d}\otimes\boldsymbol{d}). \qquad (6)$$

Substituting into the microsphere formulation gives

$$\frac{\partial \Psi_M}{\partial \mathbf{F}} = \sum_{i=1}^{k}\sum_{*\in\{s,b\}} w_i\,\rho_*\,\frac{\partial \Psi_*}{\partial \lambda_*^{d_i}}\frac{1}{\lambda_*^{d_i}}\mathbf{F}_*(\boldsymbol{d}_i\otimes\boldsymbol{d}_i), \qquad (7)$$

where $w_i$ are weights for the collocation directions $\boldsymbol{d}_i$ ($i = 1, \ldots, k$). A 21-direction integration scheme [35] is employed to evaluate the microsphere integrals.

*4.1. Thermodynamic Consistency and Frame Indifference*

The thermodynamic consistency is satisfied if the following holds

$$0 \leq \mathbf{P} : \dot{\mathbf{F}} - \dot{\Psi}_M \tag{8}$$

From continuum mechanics, we have

$$\frac{d}{dt} \ln(\det(\mathbf{F})) = \mathbf{F}^{-T} : \dot{\mathbf{F}} \tag{9}$$

Using the incompressibility constraint (i.e., $\det(\mathbf{F}) = 1$), we derive

$$\mathbf{F}^{-T} : \dot{\mathbf{F}} = 0 \tag{10}$$

Now, if we substitute the first PK stress given by Eq. 4 into Eq. 8, the thermodynamic relationship is satisfied by being equal to zero. The strain energies of both brittle and soft networks are calculated from microsphere integration over a few directional stretches ($\lambda^{d_i}$). Since $\lambda^d = \sqrt{\boldsymbol{d}\mathbf{C}\boldsymbol{d}}$ and $\mathbf{C} = \mathbf{F}^T\mathbf{F}$, for any proper orthogonal matrix $Q$, $\lambda^d$ does not change because we have

$$(\mathbf{QF})^T(\mathbf{QF}) = \mathbf{F}^T\mathbf{Q}^T\mathbf{Q}\mathbf{F} = \mathbf{F}^T\mathbf{F} \tag{11}$$

As a result, our model is frame indifferent.

*4.2. Machine Learning for Strain Energies and Network Evolution*

Equation (7) contains four unknown functions: the decay function $\rho_s$, the evolution function $\rho_b$, and the strain energies $\Psi_s$ and $\Psi_b$. These are approximated by neural networks. To impose both polyconvexity and growth conditions, we assume the following form for either soft or brittle strain energies.

$$\Psi_s(\lambda) = \text{Softplus}(ICNN_s(\lambda)) + \lambda(a_s\lambda + b_s), \tag{12}$$

$$\Psi_b(\lambda) = \text{Softplus}(ICNN_b(\lambda)) + \lambda(a_b\lambda + b_b), \tag{13}$$

where the terms $\lambda(a_s\lambda + b_s)$ and $\lambda(a_b\lambda + b_b)$ are penalty functions introduced to assure $\Psi_s(\lambda)$, and $\Psi_b(\lambda)$ convexity at finite deformations and consequently assure

$$\lambda \to \inf \quad \text{then} \quad [\Psi_s \& \Psi_b] \to \inf \tag{14}$$

Here, ICNN stands for input convex neural networks (ICNNs) [36]; $a_*$ and $b_*$ are positive and trainable variables. Note that in the above formulation, we have applied a softplus activation function to the output of ICNNs to ensure non-negativity of their outputs. Because the softplus function is convex and increasing, the first term in Eqs. 12 and 13, which is the composition of two convex functions, is convex. By construction, the strain energies are positive and convex because they are the summation of a non-negative convex function (i.e., the first term) and a quadratic polynomial with a positive leading coefficient, which makes them convex for all values of stretches (note that $\lambda$ ranges from zero to infinity). Since convexity is a stronger condition than polyconvexity, polyconvexity is automatically satisfied by the convexity of strain energies. Furthermore, as $\lambda$ tends to infinity, the strain energies also go to infinity because of the quadratic term with a positive leading term. Therefore, our model satisfies the growth condition as well as polyconvexity.

The network contributions evolve monotonically with time: $\rho_b$ increases, while $\rho_s$ decreases. To represent this behavior efficiently, we model $\rho_b$ directly with a neural network, and introduce an auxiliary function $\hat{\rho}_b := 1 - \rho_s$, which increases monotonically and is approximated by a second neural network of identical architecture. The increasing behaviour of $\hat{\rho}_b$ and $\rho_b$ is enforced by soft constraints, as explained in Section 4.5. Both $\hat{\rho}_b$ and $\rho_b$ take time, temperature, and dose rate as inputs. Figure 8 illustrates the architecture of temporal networks for $\hat{\rho}_b$ and $\rho_b$.

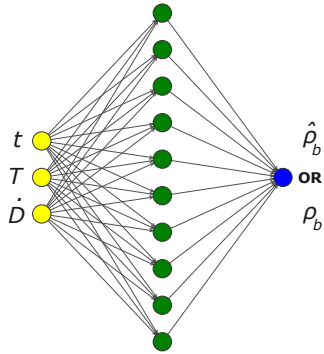

Figure 8: Neural network architecture for temporal components, namely, $\hat{\rho}_b$ and $\rho_b$

*4.3. Neural Network Hyperparameters*

This section summarizes the hyperparameters adopted for the temporal and strain energy neural networks. For the temporal components ($\hat{\rho}_b$ and $\rho_b$), shallow neural networks with a single hidden layer of 20 units were used, employing the sigmoid activation function to capture monotonic growth. For the stretch-dependent components ($\Psi_s$ and $\Psi_b$), deeper architectures were employed, consisting of two hidden layers with 5 units each, and softplus activation functions, along with a non-negativity constraint, leading to ICNNs. All networks were trained using the Adam optimizer with learning rates of 0.001.

*4.4. Loss Function*

The experimental dataset is denoted as $[\hat{\lambda}_l, \hat{\mathbf{P}}_l]$, where $l = 1, 2, \ldots, n_{\text{total}}$. The training objective is to minimize the mean-squared error between predicted and measured stresses:

$$\mathcal{L}_{\text{total}} = \frac{1}{n_{\text{total}}} \sum_{l=1}^{n_{\text{total}}} \left\| \mathbf{P}(\hat{\mathbf{F}}_l) - \hat{\mathbf{P}}_l \right\|^2, \qquad (15)$$

where $\mathbf{P}$ is computed from the constitutive model [Eqs. (4)–(7)], and $\hat{\mathbf{F}}_l$ is the deformation gradient associated with the $l$th experimental stretch. Because only uniaxial loading data are available, the deformation gradient takes the form

$$\hat{\mathbf{F}} = \begin{pmatrix} \hat{\lambda} & 0 & 0 \\ 0 & \hat{\lambda}^{-1/2} & 0 \\ 0 & 0 & \hat{\lambda}^{-1/2} \end{pmatrix}. \qquad (16)$$

At this stage, the loss function contains only the data-fitting term. In subsequent sections, we extend it with soft-constraint penalties to encode physical principles such as monotonic evolution.

*4.5. Constraint Implementation*

Hard and soft constraints [37] are essential components of data-driven constitutive modeling. Hard constraints are embedded directly in the neural network architecture, guaranteeing they are always satisfied (e.g., enforcing non-negativity of weights and biases). In contrast, soft constraints are applied through the loss function and may not be strictly satisfied during training, but they guide the model toward physically consistent solutions.

For the evolution functions $\hat{\rho}_b$ and $\rho_b$, the following constraints are imposed:

$$
\begin{aligned}
&\text{(a)} \quad 0 \leq \rho_b \leq 1, \\
&\text{(b)} \quad \rho_b(0, T, \dot{D}) = 0, \\
&\text{(c)} \quad 0 \leq \dot{\rho}_b(t, T, \dot{D}),
\end{aligned}
\tag{17}
$$

where the overdot denotes the time derivative. These have clear physical interpretations:

- $\rho_b$ represents the fraction of the brittle network; its value must remain bounded between 0 and 1.

- At the onset of aging, no brittle network exists.

- The brittle fraction must grow monotonically with time.

To enforce these conditions, $\rho_b$ is represented by a shallow neural network with a sigmoid activation:

$$
\rho_b(t, T, \dot{D}) = \sum_{i=1}^{m} \alpha_i \, \sigma_1(\beta_i t + \gamma_i T + \delta_i \dot{D} + \mu_i),
\tag{18}
$$

where $\alpha_i$, $\beta_i$, $\gamma_i$, $\delta_i$, and $\mu_i$ are trainable parameters. No bias is included at the output layer. A hard constraint is also imposed on the network parameters:

$$
\sum_{i=1}^{m} \alpha_i \leq 1, \qquad \alpha_i \geq 0.
\tag{19}
$$

Because the sigmoid activation satisfies $0 < \sigma_1 < 1$, Eqs. (18)–(19) ensure that condition (a) in Eq. (17) is always satisfied. Condition (b) can be imposed by multiplying $\rho_b$ with a Heaviside function, while condition (c) is imposed as a soft penalty term during training across the $(t, T, \dot{D})$ domain. In addition, the partitioning constraint of Eq. (2) is enforced by soft penalties.

Since introducing the generalized case $\phi := \rho_s + \rho_b < 1$ adds a *fading effect*, it accounts for irreversible mass loss and chain detachment from the load-bearing network. In practice, if mass loss is negligible, then $\rho_s + \rho_b = 1$ can be assumed, and only one network is required to describe the temporal evolution (e.g., $\rho_b$), with $\rho_s$ obtained by complementarity. For training efficiency, $\Psi_s$ is pre-trained using unaged datasets, since only the soft network exists at $t = 0$. Subsequently, $\Psi_b$ and $\rho_b$ are trained jointly using aged data.

*Loss Function Decomposition.* To integrate all constraints, the total loss is decomposed as

$$
\mathcal{L}_{\text{total}} = \mathcal{L}_{\text{data}} + \mathcal{L}_{\text{physics}} + \mathcal{L}_{\text{reg}}, \qquad \mathcal{L}_{\text{data}} = \frac{1}{n_{\text{total}}} \sum_{l=1}^{n_{\text{total}}} \left\| \mathbf{P}(\hat{\mathbf{F}}_l) - \hat{\mathbf{P}}_l \right\|^2
\tag{20}
$$

where $\mathcal{L}_{\text{data}}$ is the data-fitting term (Eq. (15)), and $\mathcal{L}_{\text{physics}}$ is introduced to enforce monotonicity and partitioning as given here

$$\mathcal{L}_{\text{physics}} = \lambda_\rho \langle \max(0, -\dot{\rho}_b) \rangle + \lambda_\phi \langle \max(0, \rho_s + \rho_b - 1) \rangle \tag{21}$$

Furthermore, $\mathcal{L}_{\text{reg}}$ is a standard weight-regularization term to prevent overfitting.

$$\mathcal{L}_{\text{reg}} = \lambda_w \|\mathbf{\Theta}\|^2, \tag{22}$$

where $\mathbf{\Theta}$ is the set of trainable network parameters. The coefficients $\lambda_\rho$, $\lambda_\phi$, and $\lambda_w$ balance the contributions of each constraint.

*Hard and soft constraints.* together act to restrict the search space of admissible solutions in a way that reflects physical reality. Hard constraints, embedded directly in the network architecture (e.g., non-negativity of weights, bounded outputs), guarantee that predictions remain within physically meaningful domains at all times. Soft constraints, on the other hand, are enforced through penalty terms in the loss function. While not strictly satisfied during training, they bias the optimization trajectory toward regions of the parameter space that comply with material physics. This combined strategy prevents the network from exploring unphysical behaviors, accelerates convergence, and ensures that the learned constitutive model remains faithful to underlying physical principles while retaining the flexibility of data-driven learning.

In practice, soft constraints are easier to implement because they require no changes to the network architecture; they are simply added as penalty terms in the loss function. However, this comes at the cost of additional evaluations during training, since each constraint violation must be computed at collocation points across $(t, T, \dot{D})$, increasing computational expense. Hard constraints, by contrast, are more efficient at runtime because they are enforced directly through the architecture (e.g., bounded activations, non-negative weights), but they require careful design of network structure and parameterization. In our framework, soft constraints enter the loss through regularization terms (e.g., monotonicity and partitioning penalties), while hard constraints are encoded structurally, such as enforcing $\rho_b \in [0, 1]$ via bounded activations and non-negative output weights.

The decomposition $\mathcal{L}_{\text{total}} = \mathcal{L}_{\text{data}} + \mathcal{L}_{\text{physics}} + \mathcal{L}_{\text{reg}}$ provides a natural way to integrate constraints into the training process. Hard constraints are enforced directly in the network architecture and parameterization (e.g., bounded activations ensuring $0 \leq \rho_b \leq 1$, non-negative weights $\alpha_i \geq 0$, or shape-function partition $\hat{\rho}_b + \rho_s = 1$. Here, $\hat{\rho}_b$ represents an ideal form of $\rho_b$ when there is no loss in the active polymer chains, and thus can be calculated as $\hat{\rho}_b = 1 - \rho_s$. Since both shape functions are derived based on the assumption of dual networks, they do not need to be included explicitly in $\mathcal{L}_{\text{total}}$, but they reduce the admissible search space of solutions.

Soft constraints, in contrast, are encoded through $\mathcal{L}_{\text{physics}}$. For example, monotonicity of $\rho_b$ and the partitioning constraint $\rho_s + \rho_b \leq 1$. These penalties bias the optimization trajectory toward physically admissible solutions without altering the architecture.

Finally, $\mathcal{L}_{\text{reg}}$ includes conventional weight-regularization terms (e.g., $\ell_2$ penalties) to control overfitting. Together, hard constraints shrink the hypothesis space, while soft constraints shape the optimization landscape through $\mathcal{L}_{\text{physics}}$. Figure 9 demonstrates the steps involved in calculating the total loss function and optimizing that function.

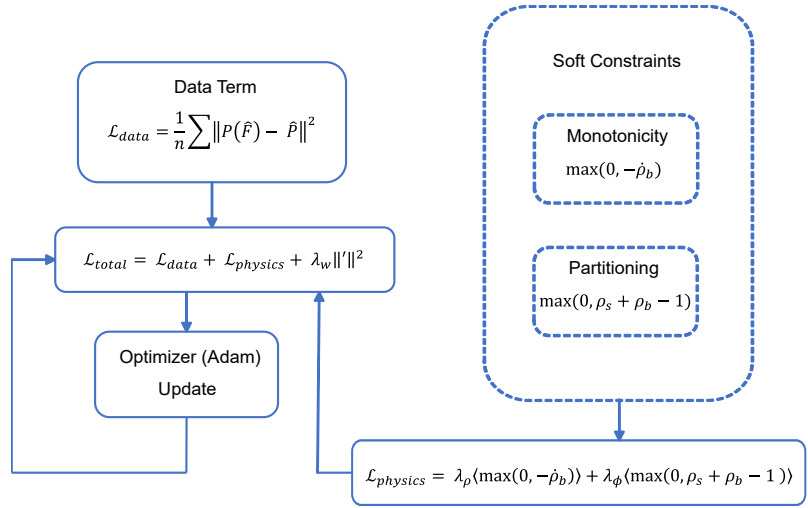

Figure 9: Flowchart of the total loss evaluation

## 4.6. Amended Loss Function

In this section, we elaborate on the implementation of soft constraints by modifying the total loss function. For illustration, we assume the material to be silicone rubber (SR); the procedure is identical for other materials. For $\hat{\rho}_b$ and $\rho_b$, we generate a collocation domain by selecting

$$
\begin{aligned}
&t \text{ time} \in \{25\,\text{h},\ 50\,\text{h},\ 75\,\text{h},\ 100\,\text{h}\}, \\
&T \text{ temperature} \in \{50°\text{C},\ 100°\text{C},\ 150°\text{C},\ 200°\text{C},\ 250°\text{C}\}, \\
&\dot{D} \text{ dose rate} \in \{0,\ 1\ \text{kGy/h}\},
\end{aligned}
\quad (23)
$$

and denote generic collocation points by $\eta_{ijk}$, where $i, j, k$ index the chosen values. The range and count of collocation points may be adjusted per material. The amended total loss is expressed as

$$\mathcal{L}_{\text{total}} = \mathcal{L}_{\text{prior}} + \sum_{i=1}^{K} \sum_{j=1}^{M} \sum_{k=1}^{N} \left[ \max(0, -\dot{\rho}_b(\eta_{ijk}))^2 + \max(0, -\ddot{\rho}_b(\eta_{ijk}))^2 + \max(0, \phi(\eta_{ijk}) - 1)^2 \right] \quad (24)$$

where $\mathcal{L}_{\text{prior}}$ corresponds to the baseline data-fitting loss [Eq. (15)], and the added penalty terms enforce monotonicity and partitioning constraints at collocation points. For SR, $\mathcal{L}_{\text{prior}}$ includes the unaged data, two data points at 25 h aging, and one data point at 50 h aging. A schematic of the training and prediction workflow is shown in Figure 10.

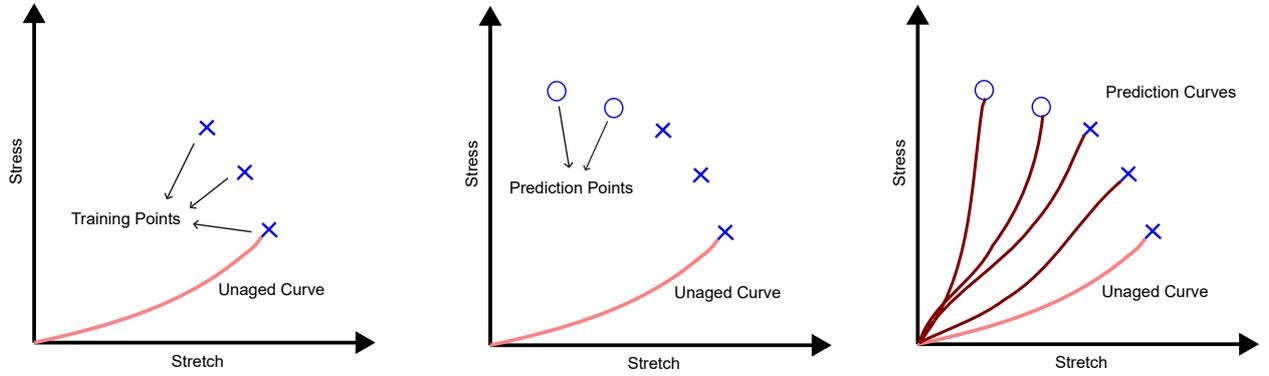

Figure 10: Training and prediction workflow: initial training on unaged data with selected failure points, followed by predictions across other thermal and radiation conditions.

## 5. Failure Prediction

Elongation at break (EAB) is central to predicting elastomer failure. Once an EAB model is identified, the stress and stretch at failure follow from combining the constitutive law with the EAB relation. To describe thermo–radiative degradation, Celina et al. [13] proposed empirical models ranging from basic to comprehensive forms. By combining their second and sixth models, the adopted time-to-equivalent-damage (TED) model [13] can be reformulated as

$$\text{TED} = \frac{\ln(\epsilon_0/\epsilon)}{[\underbrace{\tau_1 \exp(\frac{E_1}{R}(\frac{1}{T_{\text{ref}}} - \frac{1}{T}))}_{\text{thermal contribution}} + \underbrace{\tau_2 \dot{D}^x \exp(\frac{E_1(1-x)}{R}(\frac{1}{T_{\text{ref}}} - \frac{1}{T}))}_{\text{radiation contribution}}](\frac{t}{t_0})^{\alpha} g(T, \dot{D})}. \quad (25)$$

Here, TED denotes the time to equivalent damage; $\epsilon$ and $\epsilon_0$ are the aged and unaged EAB, respectively; $R$ is the universal gas constant; $t_0 = 0.01$ h; $T_{\text{ref}} = 28°$C; and $\tau_1, \tau_2, E_1, \alpha, x$ are fit parameters. The function $g(T, \dot{D}) \geq 1$ encodes thermal–radiation synergism (with $g(T, 0) = 1$ in the absence of radiation). The parameters $\tau_1, E_1, \alpha$ primarily describe thermal aging kinetics, while $\tau_2, x$ capture the radiation-specific degradation. The function $g(T, \dot{D}) \geq 1$ encodes the degree of synergism between thermal and radiative processes.

Parameter identification proceeds in two stages when data permit: (i) using thermal-only data to estimate $\tau_1, E_1, \alpha$ via nonlinear least squares; (ii) estimating $\tau_2$, $x$, and $g(T, \dot{D})$ from combined thermal–radiation data. When simultaneous data are sparse, all parameters and $g$ are identified jointly. Defining $h(T, \dot{D}) = 1/g(T, \dot{D}) \in (0, 1]$, we approximate $h$ with a neural network of two inputs and a single output (two hidden layers, 20 units each), and impose

$$h(T, 0) = 1, \qquad 0 \leq h(T, \dot{D}) \leq 1. \tag{26}$$

Nonnegativity and boundedness are enforced architecturally via a sigmoid output; the identity at zero dose is promoted by soft penalties. The EAB training objective is

$$\mathcal{L}_{\text{eab}} = \mathcal{L}_{\text{ted}} + \mathcal{L}_1 + \mathcal{L}_2, \tag{27}$$

where $\mathcal{L}_{\text{ted}}$ is the residual of Eq. (25) evaluated over the dataset, and

$$\mathcal{L}_1 = \frac{1}{N} \sum_{i=1}^{N} (h(T_i, 0) - 1)^2, \qquad \mathcal{L}_2 = \frac{1}{NM} \sum_{i=1}^{N} \sum_{j=1}^{M} \max(h(T_i, \dot{D}_j) - 1, 0)^2, \tag{28}$$

with $\{T_i\}_{i=1}^{N}$ and $\{\dot{D}_j\}_{j=1}^{M}$ forming a grid in temperature–dose-rate space. After training the EAB model, failure stress is predicted by evaluating the constitutive response at the failure stretch given by the learned EAB relation.

*Sequential vs. Simultaneous Thermal–Radiation Damage* . An important distinction in lifetime prediction is whether aging stressors act sequentially or simultaneously. In sequential aging, thermal and radiation exposures are treated as additive processes. Within our framework, this corresponds to evaluating the TED model in Eq. (25) with $g(T, \dot{D}) = 1$, such that the thermal and radiation contributions act independently and the total damage is simply the sum of their effects. In this case, $\rho_s$ and $\rho_b$ evolve under independent drivers, and their combined action is captured by superposition across separate intervals of time.

In contrast, simultaneous exposure leads to synergistic interactions where thermal energy accelerates the mobility of radiation-induced free radicals, thereby amplifying crosslinking, chain scission, and oxidative reactions. This nonlinear coupling is naturally represented in our formulation by $g(T, \dot{D}) > 1$, which increases the effective degradation rate beyond what would be predicted from independent thermal or radiation terms. The fading function $\phi(t, T, \dot{D})$ further incorporates the possibility of irreversible mass loss, ensuring that the network representation captures permanent damage.

Thus, the dual-network PINN framework distinguishes sequential versus simultaneous aging by the presence (or absence) of the synergy term $g(T, \dot{D})$ in Eq. (25), and by whether $\phi < 1$ is triggered. Sequential exposures map to additive kinetics, while simultaneous exposures require nonlinear coupling terms and stricter constraint enforcement to ensure physical fidelity.

## 6. Validation

To validate the proposed model, we compare simulation predictions against published experimental datasets for three elastomeric materials exposed to combined thermal and radiation aging. The validation tests the ability of the framework to predict long-term stress–strain behavior and elongation-at-break (EAB) using only unaged data and a limited number of failure points for training. For each case, the longest exposure times were intentionally excluded from training to test extrapolative accuracy. Figure 11 illustrates the declining loss curves for all three materials. In addition, Table 3 shows the error between experimental and simulation results.

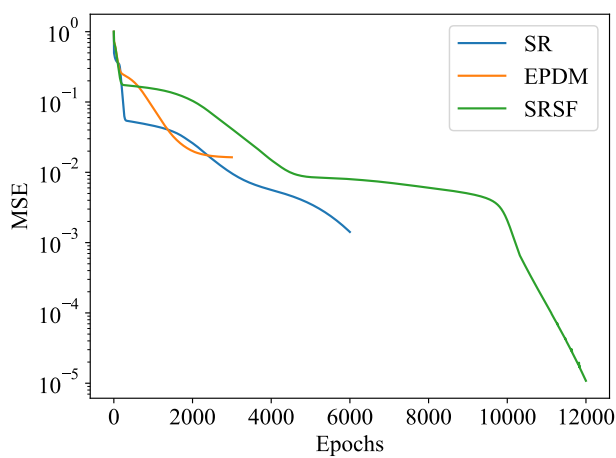

Figure 11: Loss curves for Silicone Rubber (SR), EPDM, and Silica-Reinforced Silicone Foam (SRSF).

| Material | RMSE [MPa] | MAPE [%] | EAB error at longest test [%] |
|----------|------------|----------|-------------------------------|
| SR       | 0.13       | 5.18     | 8.92                          |
| EPDM     | 0.27       | 7.6      | 12.84                         |
| SRSF     | 0.01       | 3.19     | 5.75                          |

Table 3: Quantitative validation metrics comparing model predictions with experimental datasets.

*Silicone Rubber (SR).* Shimada et al. [11] investigated silicone rubber cables for nuclear power plants under thermal–radiation aging. Their work showed that SR exhibits significant stiffening and EAB reduction with increasing dose and temperature, with degradation attributed to chain scission, crosslinking, and oxidative processes. In our study, SR was tested at 205°C and 1 kGy/h dose rate.

Two models were trained. The first included unaged data, the failure point at 25 h, one intermediate point at 25 h, and the failure point at 50 h. The second excluded the intermediate point. Figures 12 and 13 compare the two models. The first model shows superior agreement with the experimental data, particularly for the 75 h prediction, which was not used in training. This highlights the importance of including even sparse intermediate points to guide the PINN toward correct stiffening trends.

Shimada et al. reported that EAB of SR decreased sharply to about 40% of its original value after 200 kGy exposure, consistent with our model's predicted failure stretch trends. Quantitative comparison yields an EAB prediction error of 8.92% at 75 h and RMSE of 0.13 MPa on the stress–strain curve.

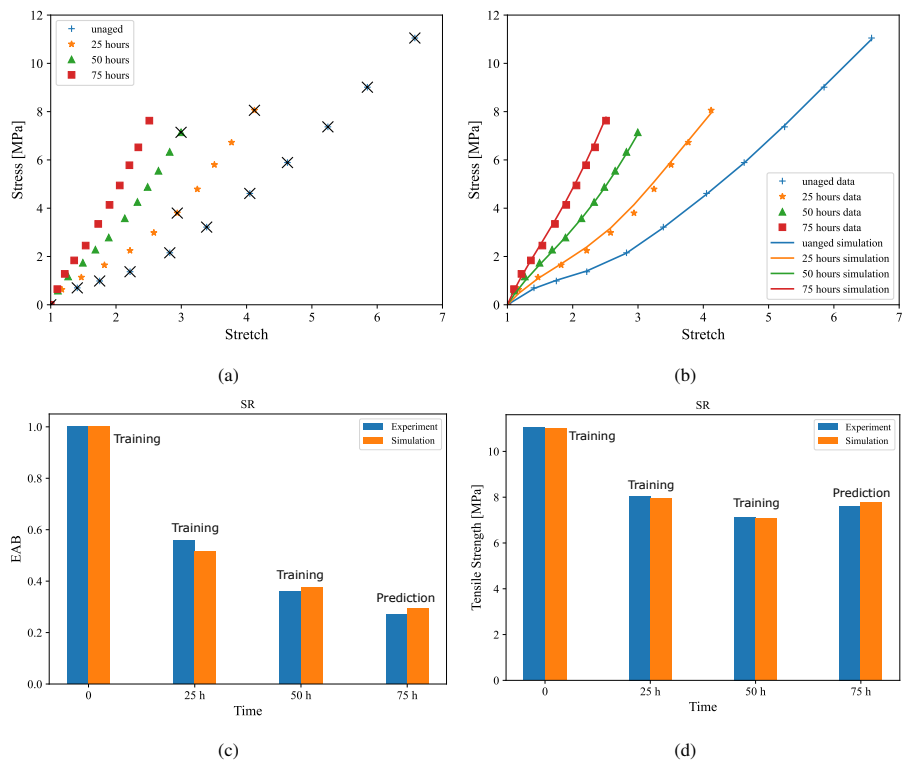

Figure 12: First SR model trained with unaged data, a 25 h failure point, a 25 h middle point, and a 50 h failure point. (a) Training dataset, where black markers denote experimental data used for training. (b) Model predictions across longer exposures, including extrapolation to 75 h not seen during training. (c) EAB experiment versus prediction. (d) Tensile Strength experiment versus prediction.

*EPDM.* EPDM is widely used in industrial cables and is known for radiation resistance compared to SR. Šarac et al. [32] compared industrial and neat EPDM under thermal–radiation aging and showed that degradation kinetics are sensitive to both formulation and environment. In our validation, EPDM was tested at multiple conditions: 70°C at 1390 Gy/h for 143 h, 55°C at 455 Gy/h for 527 h, and 25°C at 720 Gy/h for 1250 h.

Training used unaged data, two points from 70°C, and the failure point at 55°C. Figure 14 shows the predicted constitutive curves. The model successfully extrapolates to the longest aging time (1250 h), capturing the stiffening and EAB reduction trends reported by Šarac et al. They found that EAB of neat EPDM decreased by more than 50% after long-term irradiation, consistent with our predicted 48–55% reduction at 1250 h. The stress–strain RMSE was 0.27 MPa and EAB prediction error 12.84%.

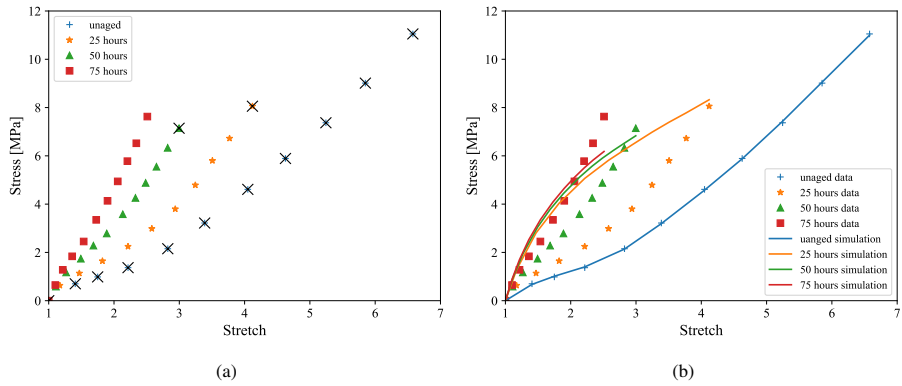

Figure 13: Second SR model trained with the same dataset as Figure 12, but excluding the intermediate 25 h point. (a) Training dataset. (b) Predictions show reduced accuracy at 75 h, highlighting the importance of sparse intermediate data for extrapolation.

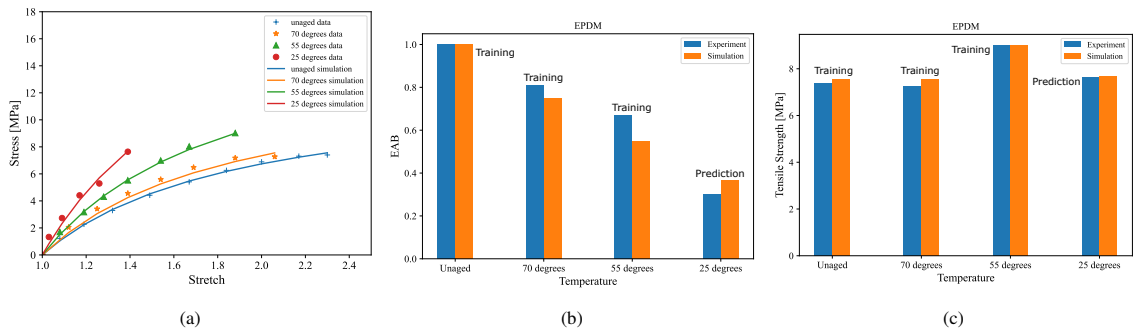

Figure 14: Predicted constitutive response of EPDM under thermal–radiation aging. Training included unaged data, two points at 70°C, and the failure point at 55°C. (a) Prediction over 1250 h. (b) EAB experiment versus prediction. (c) Tensile Strength experiment versus prediction.

*Silica-Reinforced Silicone Foam (SRSF).* SRSF is increasingly used as insulation due to its enhanced flame retardancy and mechanical stability. Fang et al. [33] studied its degradation under ~20°C and 1200 Gy/h. Their results showed rapid EAB reduction with increasing dose, with microcracking and filler–matrix debonding as dominant failure modes. In our study, doses of 100, 200, 400, and 600 kGy were considered, corresponding to 83–500 h exposures. Training included unaged data, two points at 100 kGy, and failure points at 200 and 400 kGy.

Figure 15 demonstrates accurate prediction of the 600 kGy case, which was not used in training. Fang et al. reported that EAB at 600 kGy dropped to below 30% of its original value, and our model has EAB prediction error 5.75%, again confirming agreement. The stress–strain RMSE was 0.01 MPa.

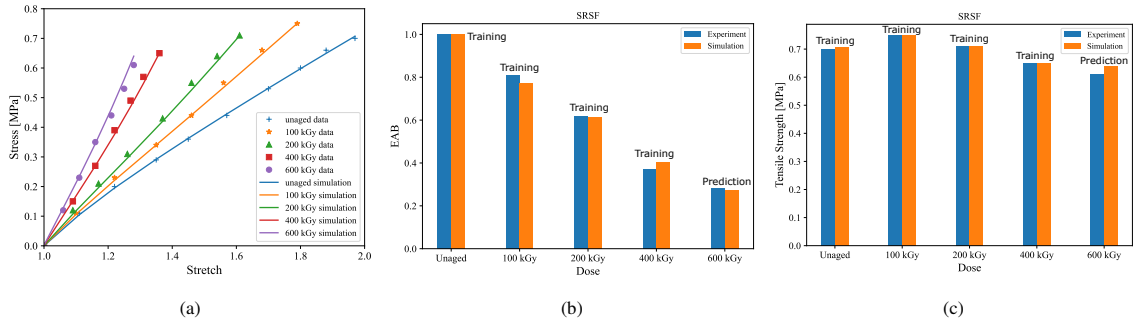

Figure 15: Predicted constitutive response of silica-reinforced silicone foam (SRSF) under simultaneous thermal–radiation exposure at 20°C and 1200 Gy/h. Training included unaged data, two points at 100 kGy, and failure points at 200 kGy and 400 kGy. (a) The model successfully extrapolates to 600 kGy (500 h) without using this condition in training, demonstrating strong predictive capability at high doses. (b) EAB experiment versus prediction. (c) Tensile Strength experiment versus prediction.

Overall, the validation confirms that the dual-network PINN framework can capture nonlinear, synergistic thermal–radiation aging mechanisms in elastomers. Future work will extend this validation to sequential exposures and lower dose rates closer to service conditions in nuclear environments.

*Uncertainty, Robustness, and Cross-Material Consistency.* Model accuracy is sensitive to the choice of training points: for SR, including a 25 h intermediate point improved 75 h predictions, underscoring the value of strategically placed data. Overall, the framework is robust to small datasets (e.g., SRSF) but extrapolation beyond tested dose/temperature regimes remains uncertain. Across materials, SR showed best agreement due to simpler degradation pathways, while EPDM exhibited larger deviations reflecting formulation complexity and varying dose rates. SRSF predictions were accurate despite fillers and foam porosity. The strategy of unaged data, failure points, and one intermediate point proved effective across cases.

## 7. Conclusion

This work demonstrates a physics-informed machine learning approach for predicting elastomer degradation in radiation–thermal environments. By combining the dual-network hypothesis with the microsphere concept, and embedding physical laws into the neural network through hard and soft constraints, the framework significantly reduces the search space and prevents non-physical solutions. Hard constraints, such as bounded network fractions, ensure structural fidelity, while soft constraints, including monotonicity, bias training toward realistic degradation pathways. Validation on silicone rubber, EPDM, and silica-reinforced silicone foam confirms that the model accurately predicts long-term stress–strain behavior and elongation-at-break using limited datasets. Including sparse intermediate data points improves extrapolative capability, and predictions remain robust across different formulations and exposure conditions. While sequential aging datasets are still required for full generalization, the proposed method highlights the value of constraint-informed learning in accelerating lifetime assessment of polymers for nuclear and high-radiation applications.

**Code Availability**

The code supporting the results of this study is available at `https://github.com/HPMroozbeh`.